\documentclass[
aps,%
12pt,%
final,%
notitlepage,%
oneside,%
onecolumn,%
nobibnotes,%
nofootinbib,%
superscriptaddress,%
noshowpacs,%
centertags,%
natbib
]{revtex4} 

\usepackage{amsmath}
\usepackage{epstopdf}
\usepackage[usenames]{color}

\begin{document}
\selectlanguage{english}

\title{The evolution of hydrocarbon dust grains in the interstellar medium and its influence on the infrared spectra of dust}

\author{\firstname{M.~S.}~\surname{Murga}}
\email[E-mail: ]{murga@inasan.ru} 
\affiliation{Institute of Astronomy, Russian Academy of Sciences,
ul. Pyatnitskaya 48, Moscow, 119017 Russia}

\author{\firstname{S.~A.}~\surname{Khoperskov}}
\affiliation{Institute of Astronomy, Russian Academy of Sciences,
ul. Pyatnitskaya 48, Moscow, 119017 Russia}
\affiliation{Universit\'{a} degli Studi di Milano,
Dipartimento di Fisica, via Celoria 16, I-20133 Milano, Italy}
\affiliation{Sternberg Astronomical Institute, Lomonosov Moscow State
University, Universitetskii pr. 13, 119992 Moscow, Russia}

\author{\firstname{D.~S.}~\surname{Wiebe}}
\affiliation{Institute of Astronomy, Russian Academy of Sciences,
ul. Pyatnitskaya 48, Moscow, 119017 Russia}

\begin{abstract}
We present evolutionary calculations for the size and aromatization degree
distributions of interstellar dust grains, driven by their destruction
by radiation, collisions with gas particles, and shattering due to grain-grain
collisions. Based on these calculations we 
model dust emission spectra. The initial grain size distribution play the important role in the evolution of an ensemble of dust particles. Radiation in the considered intensity range mostly aromatizes grains.
The smallest grains are mainly destructed via sputtering by collisions 
with gas particles. There are no grains smaller than 20~\AA\ in the medium at relative gas-dust velocities more than 50~km/s, which is typical for shocks 
in supernova remnants. The IR emission spectrum changes significantly due to the dust evolution depending on the adopted grain properties, in particular, on 
the energy of the C--C bonds ($E_0$). Aromatic bands in 
the near-IR (2--15~$\mu$m) are absent, if $E_0$ is low, even when the medium 
properties correspond to the average interstellar medium in our 
Galaxy. As in reality these bands are observed,  high $E_0$ values are more preferable. We consider
dependence of the emission intensity ratios
for various photometric bands on the medium properties. The aromatization degree of small 
grains shows up most strongly in the $I_{3.4}/I_{11.3}$ 
intensity ratio while the fraction of aromatic grains in the total dust mass influences the $I_{3.3}/I_{70+160}$
ratio.
\end{abstract}

\maketitle

\section{INTRODUCTION}

Modern ground- and space-based infrared (IR) telescopes provided us with the abundant observational material
about the composition and evolution of the dust component of the interstellar medium (ISM). One of the key steps in developing
our understanding of the dust evolution was the discovery of emission in unidentified
IR bands~[1], which are now commonly which are now commonly ascribed to the presence of so-called
polycyclic aromatic hydrocarbons (PAHs) in the ISM. These macromolecules
occupy an intermediate position between molecules and grains~[2, 3].
While PAHs do possess the emission properties necessary to explain
the observed bands, it has not yet been well established which
form they have in the ISM. In particular, it is not clear whether
aromatic compounds exist as an independent component of the ISM or
are incorporated into grains, having more diverse chemical structure~[4, 5].
In spite of this uncertainty, the abbreviation PAH is widely used to
denote the particles which are the sources of these aromatic bands.

Significant contribution into the study of the nature of aromatic bands
can be made by investigating the relationship between their properties and the
evolution of dust particles under the influence of ultraviolet (UV) radiation, 
cosmic rays, and other factors. In particular, different distributions of
the emission of aromatic dust grains at 8~$\mu$m and of other large
grains at 24~$\mu$m in so-called ``infrared bubbles''~[6, 7]
associated with regions of ionized hydrogen could provide evidence for the 
efficient destruction of PAHs by radiation from massive stars~[8].

In the interpretation of observations, value of the mass fraction of PAHs in the total dust mass,
$q_{\textrm{PAH}}$, is often used. This value is a measure of the relative contribution of
PAHs to the dust component of the ISM. Observational data in both near and far IR [9] are required
to estimate this parameter. Generally, observations with the \emph{Spitzer} and \emph{Herschel} space 
telescopes are used.  If no \emph{Herschel} data are available for a particular object, 
the ratio of the fluxes in the \emph{Spitzer} 8 and 24~$\mu$m photometric bands 
($F_8/F_{24}$) can be used as an indicator of the PAH content~[10, 11]. Information about the
nature of PAHs (or, more generally, on the nature of the sources of aromatic bands; e.g.,~[12])
can be hidden in the dependence of $F_8/F_{24}$ on the metallicity of a galaxy or an individual ionized hydrogen complex. 
It was noted in the works of Wiebe et al. [13] and Khramtsova et al. [14] that both this ratio and its
characteristic time variations depend on the metallicity in star-forming regions
(SFRs). Specifically, $F_8/F_{24}$ decreases with time in SFRs with 
solar or higher metallicities, as is expected if PAHs are destroyed by the UV
radiation of young stars more efficiently than are larger dust grains.
However, the opposite tendency is observed in SFRs with lower metallicity
--- $F_8/F_{24}$ {\emph{grows}} with time, as if the relative amount of PAHs
were increasing. There is no correlation between the PAH content and age for
SFRs of intermediate metallicity.

These results indicate that the evolution of PAHs in SFRs is a more complicated process
and it includes several processes, with  UV destruction being only one of them.
As one option, a model for cosmic dust was presented by Jones et al. [15]
(further J13) that takes into account the possible evolution of the optical
properties of grains in a SFR. In this model, the sources of emission in
IR emission bands ascribed to PAHs could be hydrocarbon grains with an
initially aliphatic structure that becomes aromatized under the action of UV
radiation.

In~[16] (further Paper~I), we presented a model for the evolution of an
ensemble of hydrocarbon grains that includes aromatization of hydrocarbon
particles, along with the more standard processes of destruction by
UV radiation and energetic gas particles and shattering in grain-grain collisions. We continue this work in our current study, and present
examples of our modeling of the evolution of an ensemble of dust grains
in the ISM and corresponding variations in their emission spectra.

\section{MODEL FOR THE EVOLUTION AND METHODS FOR COMPUTING THE 
DUST EMISSION SPECTRA}

The model we used in our study is described in detail in Paper~1, and we
will only list its main characteristics here. It is assumed in the model that, at
an initial time (corresponding, for example, to the formation of dust in
evolved stars or the onset of star formation in a molecular cloud), the
ensemble of dust particles is characterized by specific distributions of the 
grain sizes and aromatization degrees, with the latter described by the width 
of the energy band gap of the grain material $E_{\textrm{gap}}$. During the
evolution of the ensemble, these distributions change due to aromatization and 
dissociation via the absorption of UV photons and collisions with energetic 
gas particles (non-thermal and thermal ions and electrons) and shattering 
during collisions between grains. Our goal here mainly is to investigate the
evolution of grains in an ISM with a relatively modest density, where
the growth of grains as a result of accretion of gas and coagulation is 
not efficient, so that these processes can be neglected in this work. 
We plan to include such processes in future studies.

Grain sizes are redistributed because of shattering, in particular, 
increasing the fraction of small grains. Under certain conditions, 
shattering could be the main origin for small dust grains in the ISM~[17]. In the absence of shattering, 
small grains are rapidly destroyed by photons, ions, and electrons, so that 
the ISM would be left without any small grains. In the context 
of our study, this may be the key process leading to the presence of a 
noticeable amount of aromatic compounds in the ISM, which are responsible 
for the observed IR emission bands. Even if fragments arising from the 
destruction of large grains are mainly hydrogenated and initially have an 
aliphatic structure, they are rapidly aromatized by the UV radiation.

If conditions are such that the rate of aromatization of small grains is
comparable to the rate of their formation, we will observe only aromatic
grains in the ISM. If the shattering rate for aliphatic grains exceeds
the aromatization rate of the fragments, both aliphatic and aromatic dust
grains will be observed. Since astrophysical objects usually do not exist
under uniform conditions, the relative abundances of dust grains of various
types can change even within a single object. Interpretation of IR spectra
requires spatially resolved computations of the evolution of grains,
simultaneously modeling all important evolutionary processes under specified 
and possibly varying external conditions. We present here such computations 
for fixed conditions in the medium, but investigate the sensitivity of the 
model to changes in these conditions.

The formation and growth of carbonaceous grains in the atmospheres of
evolved stars has been described in detail in [18], however there is no
consensus about the initial structure of such grains. Theoretical
models predict that the formation of grains in envelopes of carbon
stars should mainly lead to the synthesis of aromatic particles (see [19]
and references therein). On the other hand, a comparison of aromatic and
aliphatic bands in IR spectra of young planetary nebulae shows that
grains in most of these objects have predominantly aliphatic structure~[20]. 
Before these grains penetrate into the ISM, the balance can shift
to a greater predominance of aromatic grains due to the action of the UV
radiation of the nucleus of the planetary nebula. We assume in our study that
all grains are initially aliphatic, although our model can also handle
different initial conditions.

The most important parameters of our model are the gas temperature $T$, the 
number densities of H$^{+}$, He$^{+}$, and electrons,
the radiation field, and the velocity of the relative non-thermal motion
of the gas and dust, for example, due to the propagation of shocks. We
characterize the radiation field using the parameter $U$, equal to the
radiation intensity from 912~\AA\ to 2500~\AA, expressed in units of
the corresponding radiation intensity in the solar neighborhood determined
in~[21]. In other applications our model can use different types of spectra, including
those with an appreciable contribution at wavelengths shorter than 912~\AA.

The dust is described by a two-dimensional distribution over the grain radius 
$a$ and the parameter $E_{\textrm{gap}}$, with the number of bins
being $N^{\textrm{a}}$ and $N^{\textrm{eg}}$, respectively. 
Although the smallest grains considered in the work have properties more similar to molecules than to dust particles, for
the sake of uniformity, we refer to dust particles of any size as grains. 
Of the parameters describing the evolution of dust presented in Paper~1,
we also varied the valus of $E_0$ characterizing the energy of the C--C bonds.
This energy is poorly known, and the computational results depend quite
substantially on our choice of its value, making it necessary to treat
$E_0$ as a parameter of our model.

The method used to compute the evolution of the grain distributions over
their size and $E_{\textrm{gap}}$ is described in detail in Paper~I. Here, we
also present modeling of spectra for a given ensemble of grains. The
emission coefficient $j_{\nu}$ for dust consisting of $N_{\textrm{type}}$ 
types of grain heated by a radiation field $U$ was computed using the formula
\begin{gather}
j_{\nu} = \sum\limits_{i=1}^{N_{\textrm{type}}}\int
\left(\frac{dn}{da}\right)_{i} C_{\textrm{abs}}(i,a,\nu) da
\times\\
\nonumber{}\times \int B_{\nu}(T) \left(\frac{dP}{dT}\right)_{i,
a, U}dT,
\end{gather}
where $(dP/dT)_{i, a, U}$ is the temperature distribution for grains of
type $i$ with radius $a$ in the radiation field $U$, and $(dn/da)_i$ is
the size distribution of grains of type $i$. By various grain types,
we mean hydrocarbon grains with various values of $E_{\textrm{gap}}$ and grains
with other chemical compositions, such as silicate grains, which are
included in our model for the computation of the spectra.

We take into account the fact that small grains (with radii less than
${\sim}250$~\AA) are sensitive to single photon absorption, which
heat these grains to high temperatures on short time scales, after which
they cool rapidly via IR radiation~[22, 23]. An ensemble of such grains 
in the presence of  stochastic heating cannot be characterized by a single
temperature, and we need to consider the distribution $dP/dT$, which
can be computed using the method presented in~[24, 25]. The temperature of
large grains can be found from the condition of thermal balance; the
temperature distribution of such grains is a delta function. The heat
capacity evaluation for hydrocarbon grains $C_V$ is required to
calculate their temperature distribution. It was performed as described in Paper~1. The
heat capacity of silicate grains was calculated as in~[24].

\begin{figure*}[t!]
\includegraphics[width=0.4\textwidth]{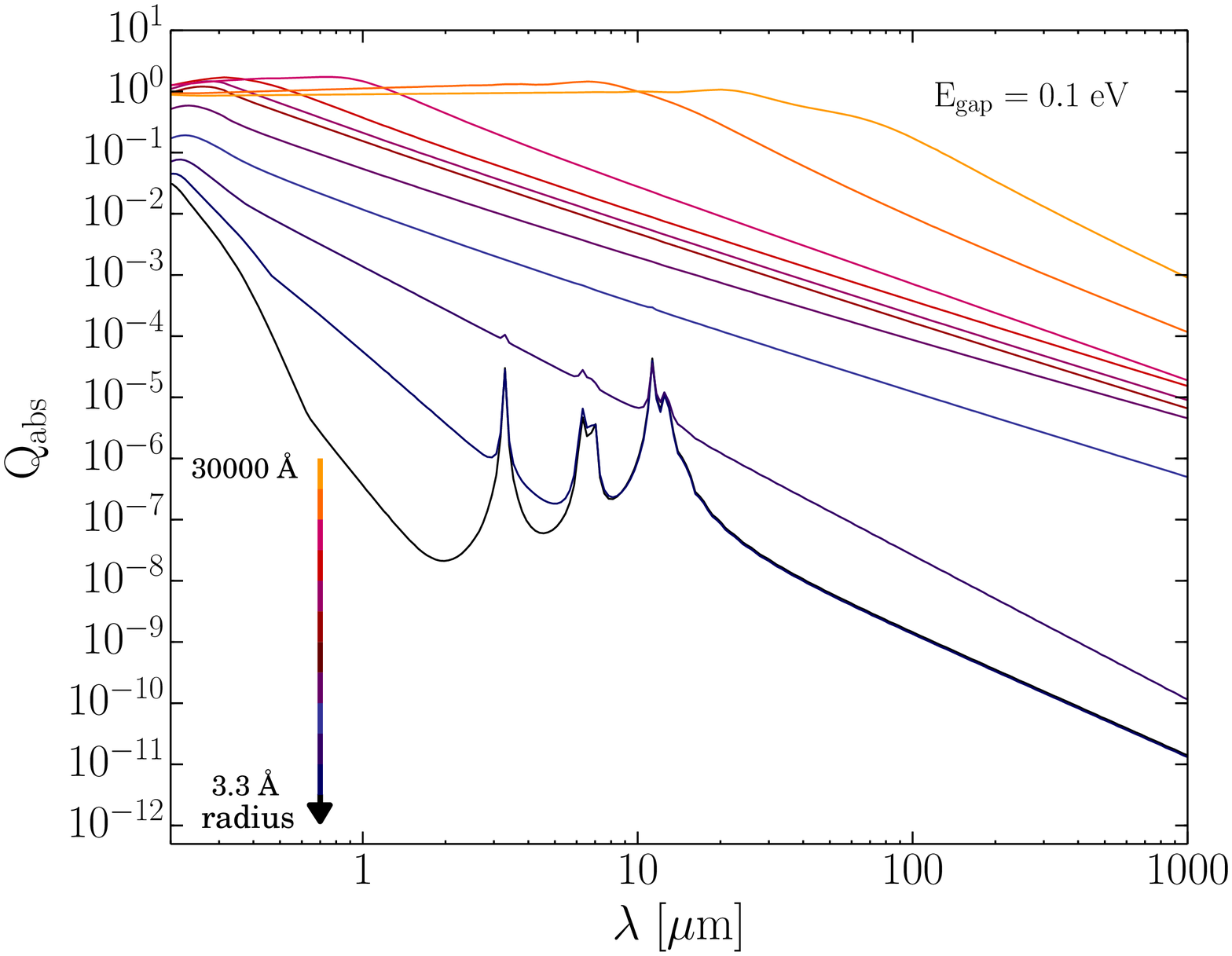}
\includegraphics[width=0.4\textwidth]{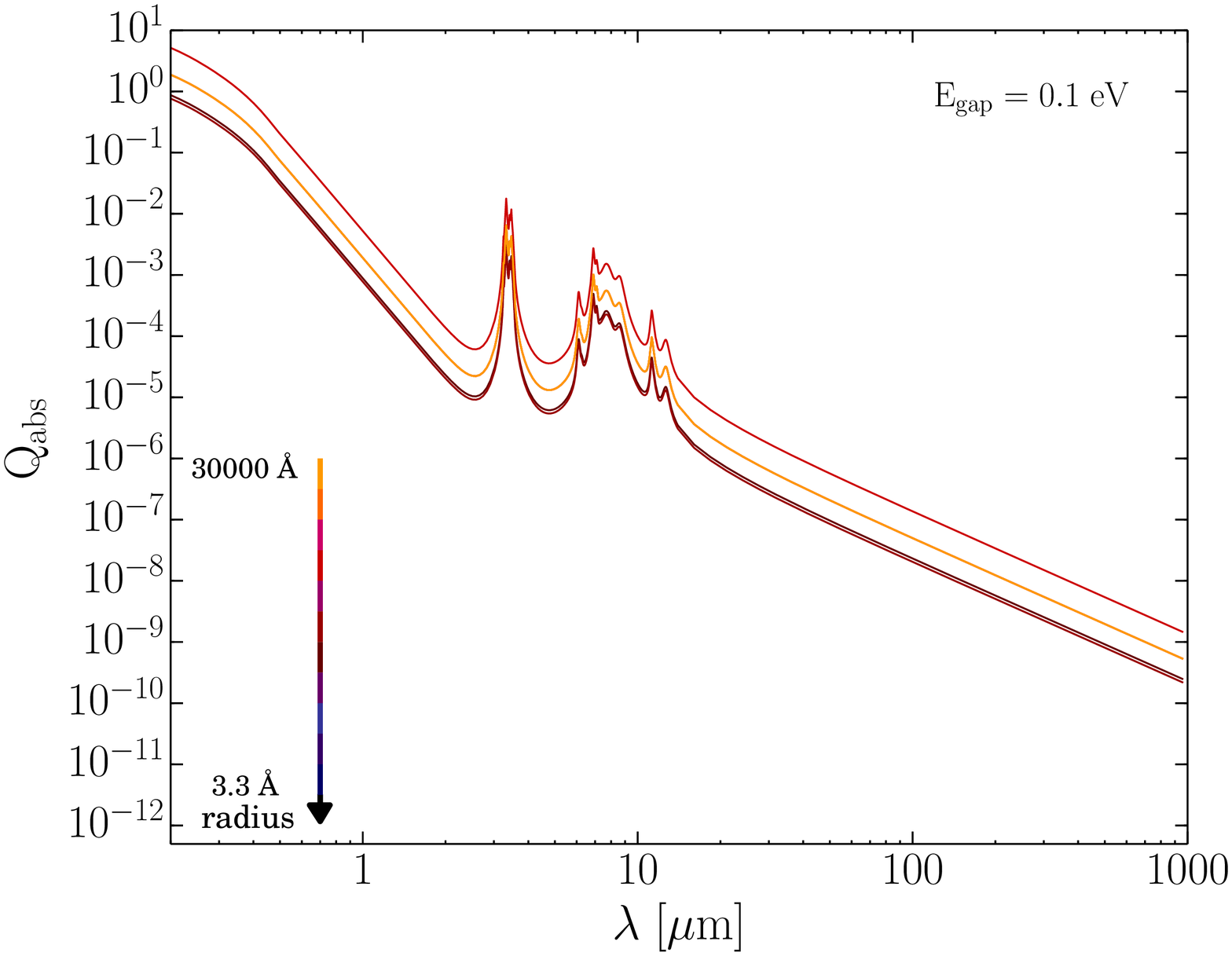}
\setcaptionmargin{5mm}
\onelinecaptionsfalse 
\caption{Wavelength dependence of the absorption efficiency $Q_{\textrm{abs}}$ 
with fixed $E_{\textrm{gap}} = 0.1$~eV for hydrocarbon grains with radii from
3.3 to 30\,000~\AA{} (the color varies from orange to black with decreasing
grain size). The left panel shows results for grains that are fully
aromatized and have radii not exceeding 200~\AA. It is assumed that only
an outer layer 200~\AA\ thick of large grains is aromatized.
The right panel shows results for grains that are fully aromatized independent
of the size. \hfill}
\end{figure*}

The optical constants (the real and imaginary parts of the refraction index
$n$ and $k$) of hydrocarbon grains with radii up to 200~\AA{} were taken 
from~[26--28] (see also~[29, 30]). We assumed that small grains are aromatized
throughout their volume, i.e., they have a uniform aromatization degree.
We assume large grains ($a>200$~\AA) to have two layers: the inner layer 
consists of hydrogenated, amorphous hydrocarbon material with
$E_{\textrm{gap}} = 2.67$~eV, while the outer layer may be restructured
by incident UV radiation, so that, in general, the optical constants for the 
surface layer may differ from those for the inner layer. We computed the
absorption efficiency $Q_{\textrm{abs}}=C_{\textrm{abs}}/\pi a^2$ for 
two-layer grains using code based on the Mie theory, described in~[31]. The
optical properties of silicate grains are taken from~[32, 33]. These grains
contribute only to the continuum emission at the considered conditions.

The left panel of Fig.~1 shows $Q_{\textrm{abs}}$ for hydrocarbon grains
of various sizes with fixed $E_{\textrm{gap}} = 0.1$~eV. The behavior of
$Q_{\textrm{abs}}$, and therefore $j_{\nu}$, at wavelengths from 2--20~$\mu$m 
depend strongly on the grain size. Bands are weak for grains with radii
of about 200~\AA{}, but become stronger with decreasing $a$. Since most 
larger grains are comprised of hydrogenated material, the characteristics
of the relatively thin aromatized surface layer are ``muffled''. For
comparison, the right panel of Fig.~1 shows $Q_{\textrm{abs}}$ for grains
of various sizes without any restriction on the aromatization depth (i.e.,
the grains are fully aromatized regardless of their size).

Figure~2 shows the absorption efficiency of grains with radii of 10~\AA{} 
and various values of $E_{\textrm{gap}}$. Predominantly aliphatic
($E_{\textrm{gap}} = 2.67$~eV, blue curves) and predominantly aromatic 
($E_{\textrm{gap}} = 0.1$~eV; orange curves) grains are characterized by peaks
in $Q_{\textrm{abs}}$ at various wavelengths or bands with appreciably
different maximum values and widths. The most substantial $Q_{\textrm{abs}}$
peaks for aromatic compounds are at 3.3, 6.3, 6.7, 11.3, and 12.7~$\mu$m,
while aliphatic compounds have their main peaks at 3.4 and 7.0~$\mu$m, with
their $Q_{\textrm{abs}}$ peaks at 11--12~$\mu$m being appreciably weaker.

\begin{figure*}[t!]
\includegraphics[width=0.7\textwidth]{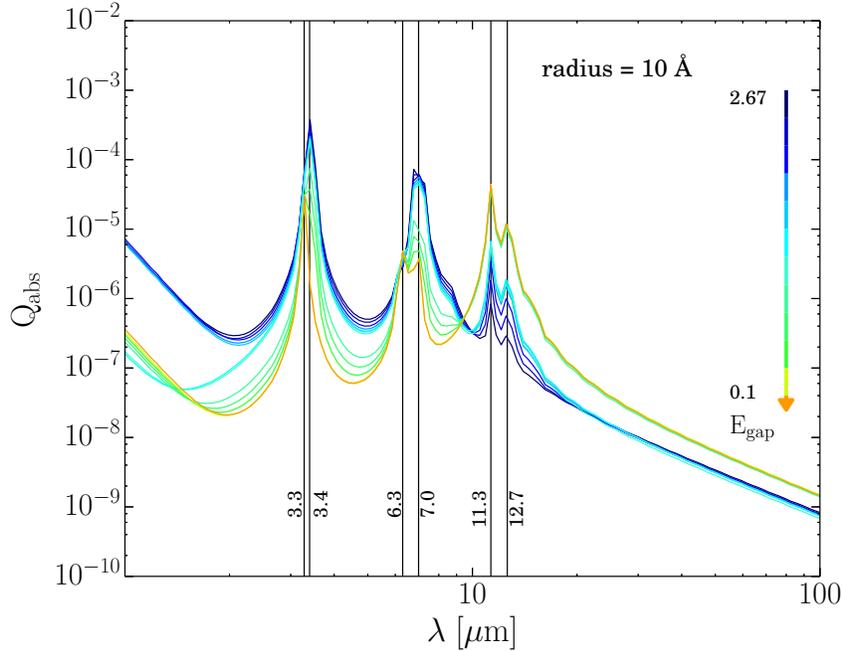}
\setcaptionmargin{5mm}
\onelinecaptionsfalse  
\caption{Wavelength dependence of the absorption efficiency 
$Q_{\textrm{abs}}$ for grains with radii of 10~\AA{} and $E_{\textrm{gap}}$ 
values from 2.67 to 0.1~eV (the color varies from orange to blue with
increasing $E_{\textrm{gap}}$). \hfill}
\end{figure*}

The ratio of the intensities of the spectral features at 3.3 and 3.4~$\mu$m
is often used to estimate the relative abundance of aromatic and aliphatic
compounds~[34--36], but obviously spectra with higher resolution are required
to separate these features. Moreover, it can be seen in
Fig.~2, the bands at 3.3 and 3.4~$\mu$m may overlap, depending on
which grain type dominates.

\section{EVOLUTION OF THE DISTRIBUTIONS OF GRAIN SIZE AND AROMATIZATON DEGREE}

In this section, we consider variations of the distributions of the 
size and aromatization degree of hydrocarbon grains. The number density
of hydrogen ions $n_{\textrm{H}^{+}}=1$~cm$^{-3}$, number density of helium
ions $n_{\textrm{He}^{+}}=0.1$~cm$^{-3}$, and temperature $T=10^4$~K of
the medium were fixed in the calculations. The parameter $U$ characterizing
the radiation field, which is the main factor influencing the aromatization
process, and the gas velocity $v_{\textrm{ion}}$, which is a determining
factor in the destruction of small dust grains, were varied. As in~[37],
the relative velocity of the grains used when modeling shattering was 
taken from~[38] for a thermally ionized medium, where the dust dynamics 
were computed for a turbulent medium using a magnetohydrodynamical model.
We used the J13 distribution proposed in [15] as the initial size distribution 
of the \textrm{a-C:H}\ grains, unless stated otherwise. This distribution
assumes that the partially aromatized hydrocarbon dust consists of small grains
with radii from 4 to 1000~\AA, whose size distribution is a power law with
an exponential tail, and large grains with a lognormal size distribution
and a characteristic radius of ${\sim}2000$~\AA. The J13 distribution also
includes silicate grains with a lognormal size distribution and a characteristic 
radius of ${\sim}2000$~\AA. We take into account the presence of silicate 
grains only when modeling spectra, and assume that the silicate grains are 
not subject to evolutionary variations. This assumption is justified because 
the destruction of silicate grains is inefficient under the conditions 
considered~[37].

\begin{figure*}[t!]
\includegraphics[width=0.4\textwidth]{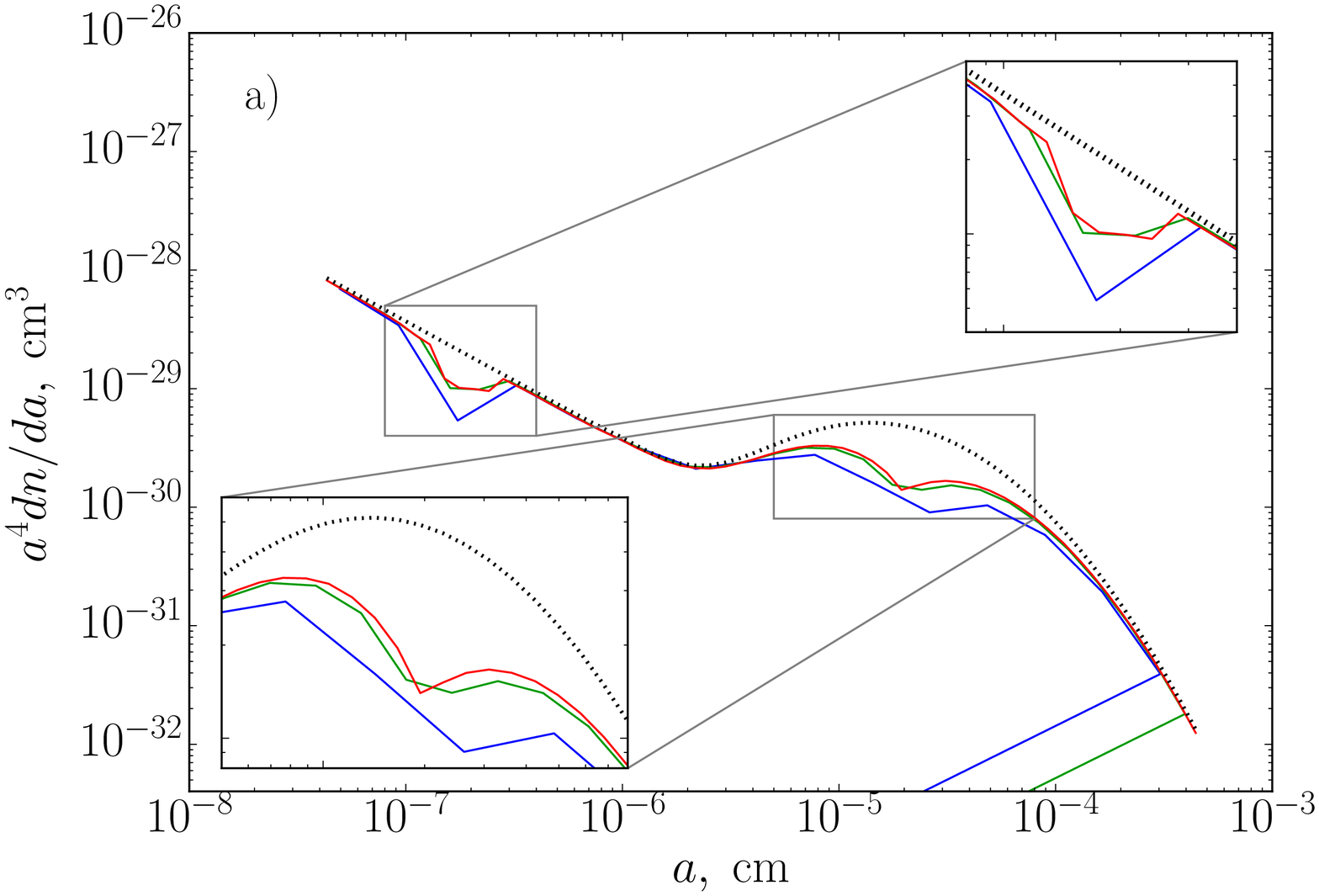}\\
\includegraphics[width=0.4\textwidth]{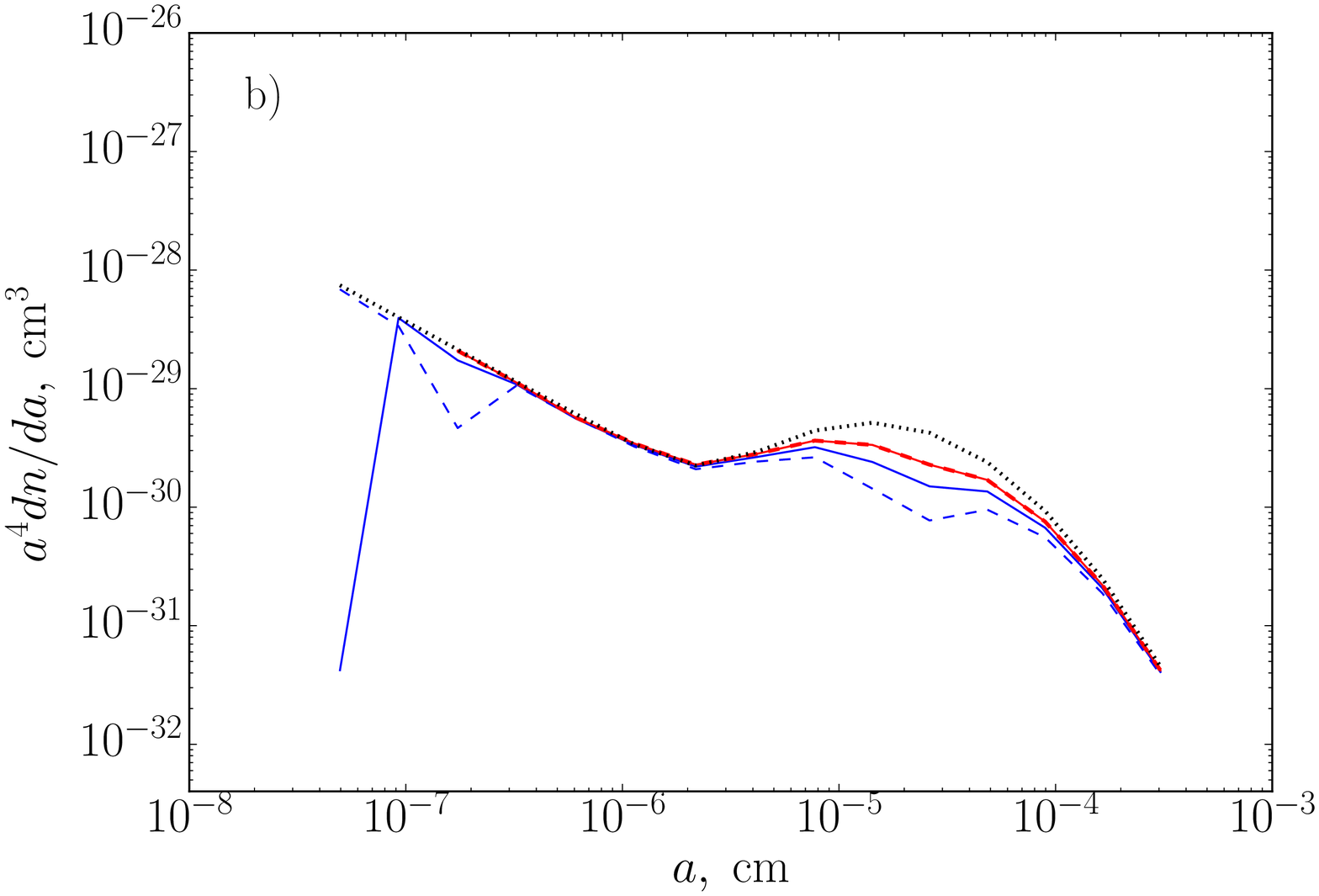}\\
\includegraphics[width=0.4\textwidth]{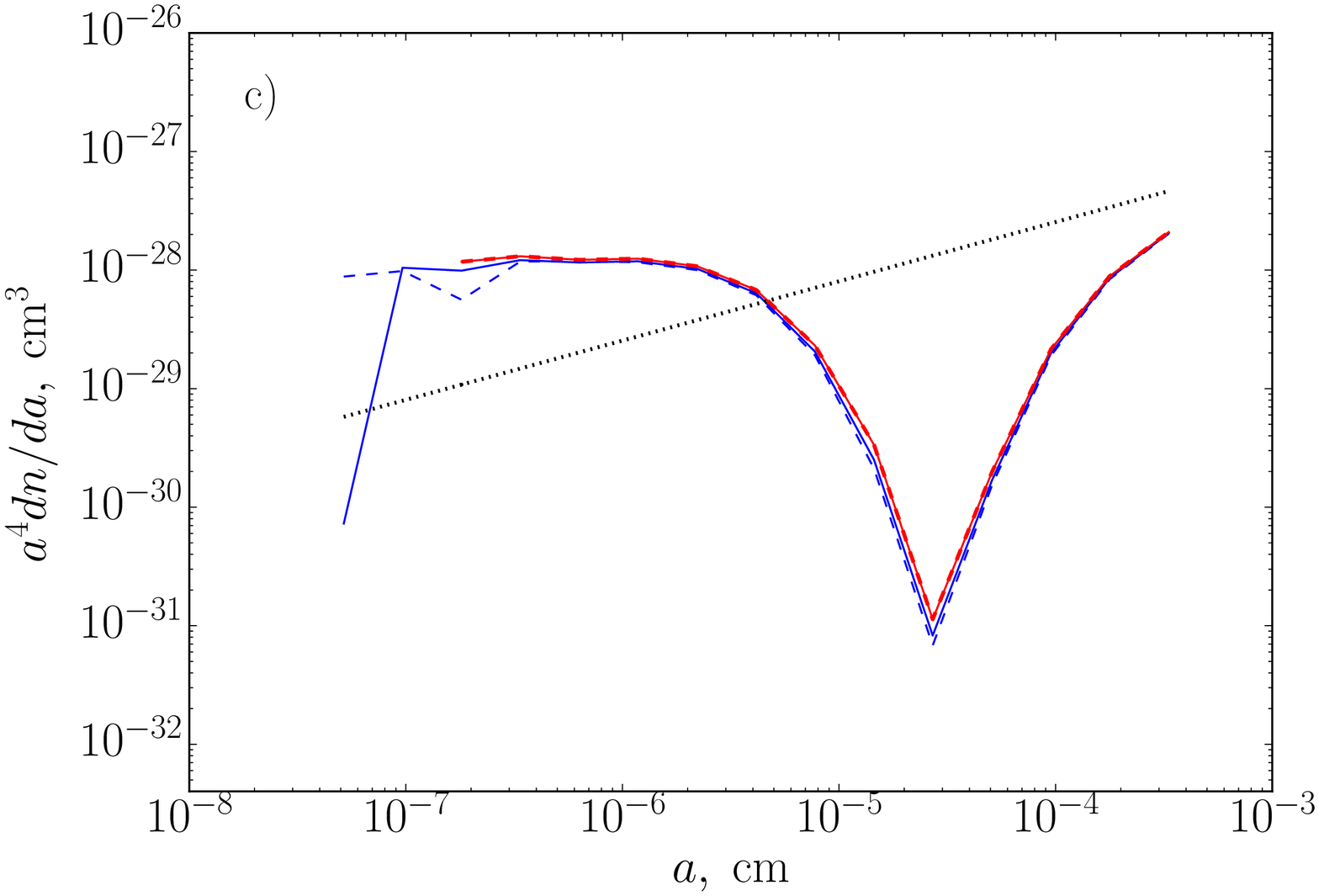}
\setcaptionmargin{5mm}
\onelinecaptionsfalse  
\caption{Size distributions of dust grains for various conditions for
computations covering one million years of evolution. (a) Red, green,
and blue curves show the results of computations for $U=1$ and 
$v_{\textrm{ion}} =10$~km/s with $N^{\textrm{a}} = 60, 30,$ and 15,
respectively. The initial distribution (J13) is shown by the dotted black
curve. (b)~Blue curves show the grain size distribution after one
million years of evolution for $U=1$ and $v_{\textrm{ion}} =10$~km/s for
$E_0=2.9$~eV (solid) and $E_0=5.0$~eV (dashed). Red curves show the
grain size distribution for $U=10^5$ and $v_{\textrm{ion}}=50$~km/s for
$E_0=2.9$~eV (solid) and $E_0=5.0$~eV (dashed). The initial distribution 
(J13) is shown by the dotted black curve. (c) Same as (b) but adopting
the MRN distribution [39] as the initial distribution. \hfill}
\end{figure*}

Figure~3 shows examples of finite grain-size distributions summed up
by the aromatization degree for several different
cases. In all panels, a dotted black curve shows the initial distribution.
To estimate the minimum number of bins required to obtain realistic
results, we considered runs with $N^{\textrm{a}}=15$, 30, and 60 and
with $E_0=5.0$~eV after one million years of evolution, shown by blue,
green, and red curves in Fig.~3a. The computations for $N^{\textrm{a}}=30$ 
and 60 are in good agreement, but reducing $N^{\textrm{a}}$ to 15 leads 
to overestimation of the shattering rates for some size bins. This is
consistent with the conclusions of Hirashita and Yan~[37], who found the
optimal number of bins for computations with shattering to be 32.
However, in addition to size binning, we also consider
several (five in this study) bins in the aromatization degree, so,
for example, for $N^{\textrm{a}}=30$ the total number of bins  is
150, leading to substantial losses in the computation speed for
a given accuracy. Therefore, we conclude that, when incorporating the model
in the computations for the real object in 1D or even more so in 3D,
it is still acceptable to use $N^{\textrm{a}}\sim15$, since the 
errors that arise due to the insufficient number of bins are smaller than
evolutionary changes in the distribution. Further,
we will present results for our computations with $N^{\textrm{a}}=15$.

\begin{figure*}[t!]
\includegraphics[width=0.8\textwidth]{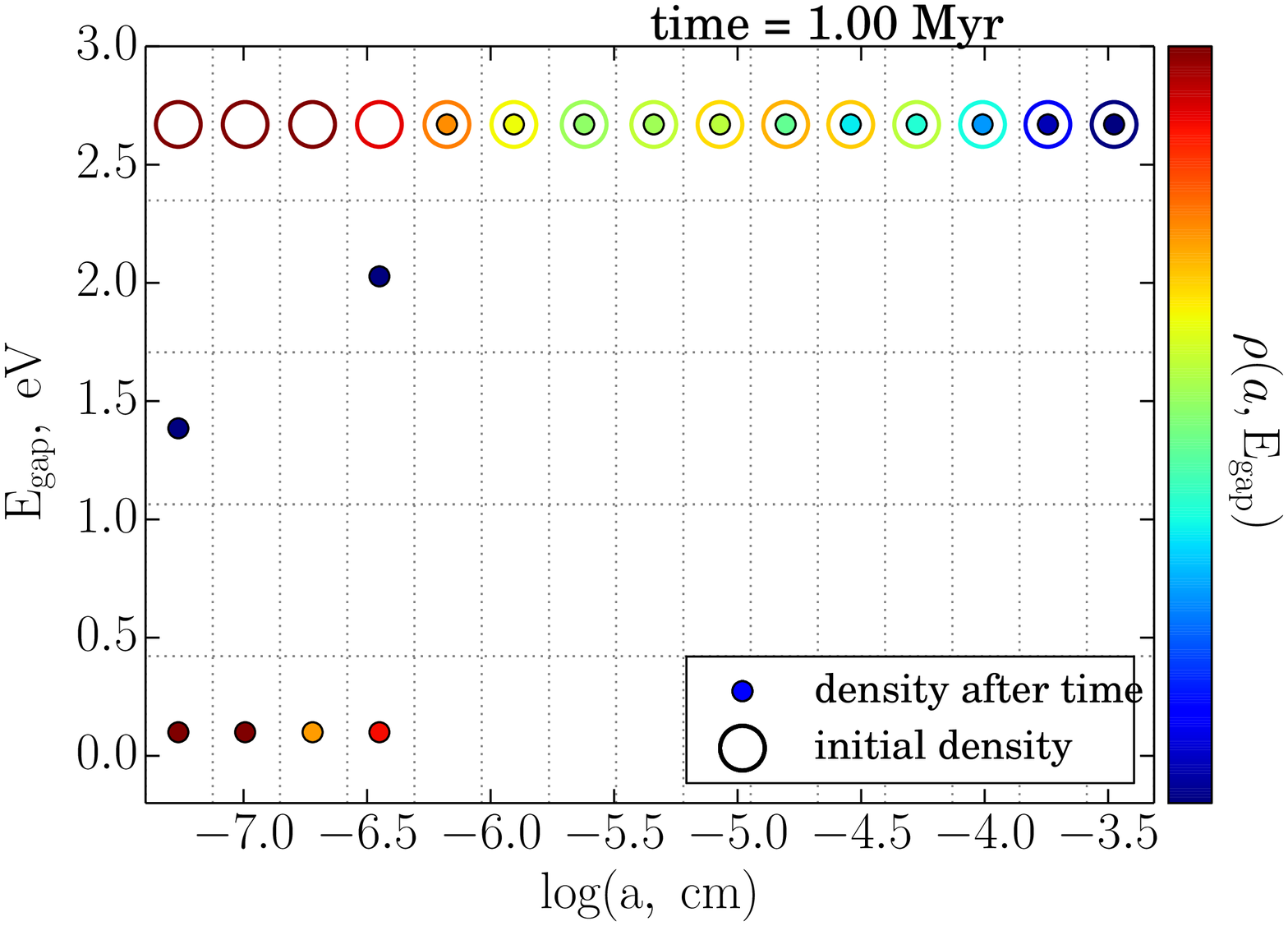}
\setcaptionmargin{5mm}
\onelinecaptionsfalse 
\caption{A $15\times5$ field separated into cells in the grain radius $a$ and
aromatization degree $E_{\textrm{gap}}$. If grains are present in the
distribution at a given time, a circle is placed in the cell for the 
corresponding parameters. Otherwise, a cell is empty. The color of the
circle corresponds to the mass density $\tilde{\rho}(m,E_{\textrm{gap}})$ 
for the grains of the given type. Grains of all sizes are unaromatized at the 
initial time; their positions in the grid are shown by empty circles.  \hfill}
\end{figure*}

Further, we carried out computations of the grain destruction for
parameters typical of the ``ordinary'' ISM ($U=1$, 
$v_{\textrm{ion}}=10$~km/s) and for ``extreme'' parameters ($U=10^5$, 
$v_{\textrm{ion}}=50$~km/s). The results are shown in Fig.~3b. The blue
curves show the grain size distribution after one million years of evolution
in an ``ordinary'' ISM for $E_0=2.9$~eV (solid) and $E_0=5.0$~eV (dashed).
At the smaller value of $E_0$, the efficient destruction leads 
to a virtual absence of grains with sizes less than $10^{-7}$~cm. A second
important result is the smoothing of the maximum corresponding to large
grains ($a\sim10^{-5}$~cm). The destruction of small grains is not as
efficient when $E_0=5.0$~eV, leading to a more substantial reduction in
the number of large grains, since the main (essentially the only) mechanism
for the destruction of grains in an ambient medium with the specified
parameters is collisions with other grains, with small grains playing an
important role. Moreover, the efficient destruction of large grains helps to
retain the high (compared to the case of $E_0=2.9$~eV) abundance of
the smallest grains.

The results of the evolution of the grain size distribution under the
``extreme'' conditions are shown by the red curves in Fig.~3b, which also
correspond to a computed time of $10^6$~yrs. Essentially the only result of
this evolution for both considered values of $E_0$ with the J13 initial
distribution is the efficient destruction of small grains and, accordingly,
a reduced destruction of large grains due to the dearth of small grains
(the solid curve for $E_0=2.9$~eV and the dashed curve for $E_0=5.0$~eV
essentially coincide).

Figure~3c shows the grain size distribution computed with an MRN distribution
[39] as the initial distribution. The MRN distribution differs from the J13 distribution
in having a smaller initial fraction of small grains. In this case,
the main factors,  changing the distribution, are the shattering of large
grains, which leads to a significant increase in the number of small grains, and
the destruction of small grains by radiation and/or collisions with gas
particles. In the distribution (Fig.~3c), this is expressed through
a sharp minimum near radii of ${\sim}3\times10^{-5}$~cm,
an increase in the number of grains with radii of $\lesssim3\times10^{-6}$~cm,
and the disappearance of grains with radii of $\lesssim10^{-7}$~cm. Since the
main evolutionary factor for the grain size distribution in the considered case is 
shattering, whose parameters we have not varied, the results for various values
of $E_0$, $U$, and $v_{\textrm{ion}}$ are all roughly the same.  

Another way of presenting results of computations is used in Figure~4. Here we show
a grid with
cells corresponding to various sizes and aromatization degrees. A cell is
occupied at a given time if teher are grains in the model with the corresponding
$a$ and $E_{\textrm{gap}}$ values. Initially, grains of all sizes
are fully hydrogenated; i.e., $E_{\textrm{gap}}=2.67$~eV. The initial 
distribution of grains in Fig.~4 is shown by large empty circles in
the upper row of the grid. Filled smaller circles correspond to the
grain distribution after one million years of evolution with $U=1$ and 
$v_{\textrm{ion}}=10$~km/s. The pattern has changed: in the size range
$\log a(\textrm{cm})<-6.3$, the dust has moved to cells with a higher
aromatization degree. The color of the circles in Fig.~4 corresponds to
the mass density in a given cell. With the specified parameters and the
J13 initial size distribution, most of the bins lose mass, but under other
conditions (for example, with the MRN initial distribution ~[39]), the amount
of dust can grow in low mass bins. When other values of $U$ are
used, small hydrogenated grains may be preserved in the grid, or, on the
contrary, large aromatized grains may appear.

Generally, grain size distribution and $E_{\textrm{gap}}$
depend on the conditions in the medium, that it, on the radiation field, temperature,
and velocities of gas and dust. The external factors influence mostly
small grains, leading to these grains being
``swept out'' of the distribution. With typical parameters of the ISM,
collisions with gas particles are a more efficient destructive factor, while
the radiation field only aromatizes small grains. It is important to remember
that, in the considered model, aromatization is a necessary step preceding
the efficient grain destruction.

The action of external factors on large grains only shows up
in the case of extreme values of the corresponding parameters. ``Softer''
values of $U$ and $v_{\textrm{ion}}$ probably facilitate the preservation
of large grains, since small grains are efficiently destroyed in this case,
so that the shattering of large grains due to collisions with small grains
becomes less efficient. We do not present here results of computations
for other densities and temperatures, but the conclusions of Paper~1 suggest
that these parameters also play an important role. Overall, the action of
external factors and interactions between grains could substantially change
the parameters of the dust population. The result of this evolution depends
appreciably on a combination of many circumstances, which can explain the
observed variations in the properties of dust grain ensembles in various
regions of the Milky Way and other galaxies. This is supported, in particular,
by observations of the extinction curves and other properties of dust in
various directions~[40, 41]. Variations in the dust parameters
can probably be explained by the variety of external conditions in which
the dust resides, which makes it impossible to build a universal model,
suitable for any object.

\begin{figure*}[t!]
\includegraphics[width=0.4\textwidth]{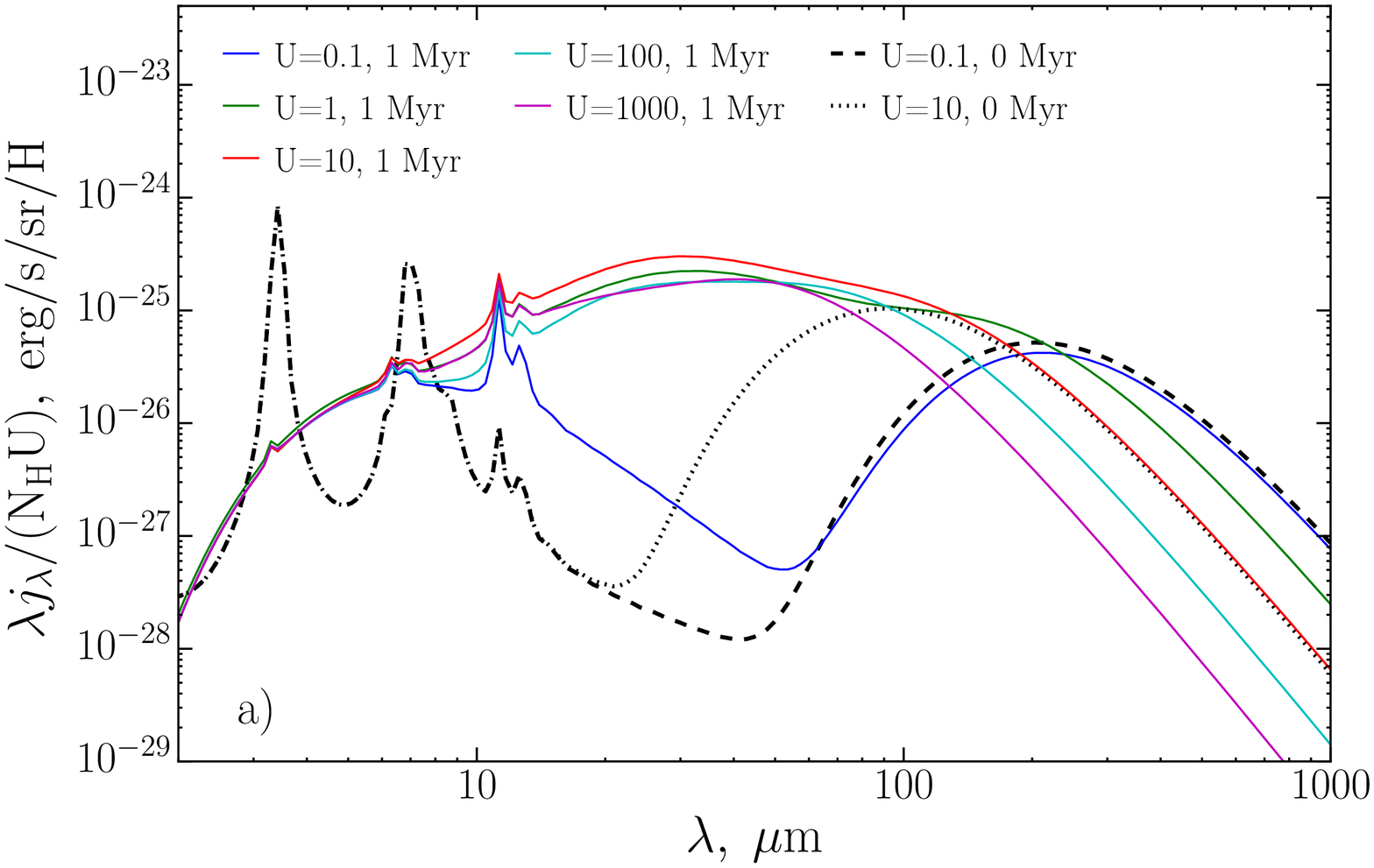}
\includegraphics[width=0.4\textwidth]{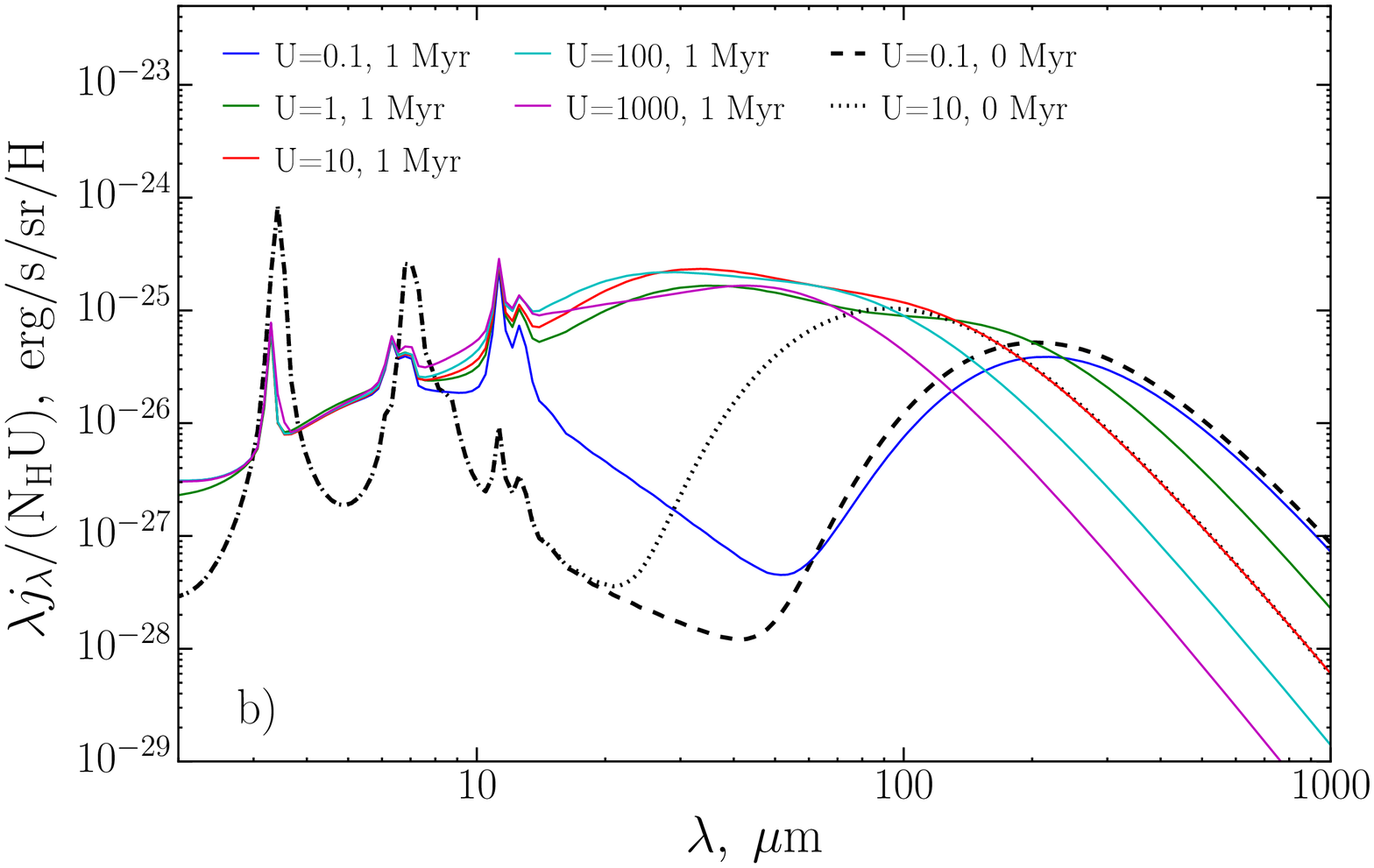}
\includegraphics[width=0.4\textwidth]{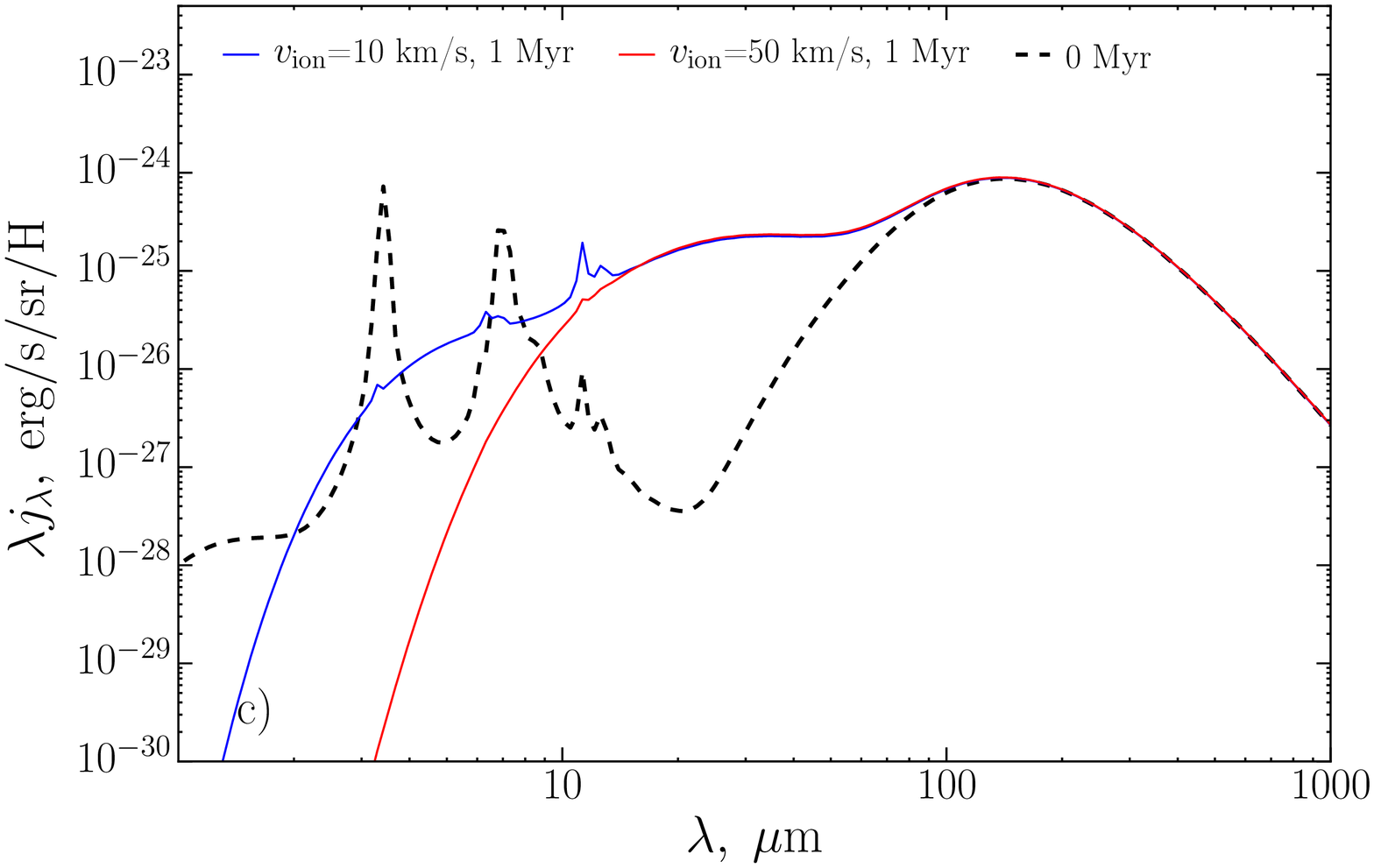}
\includegraphics[width=0.4\textwidth]{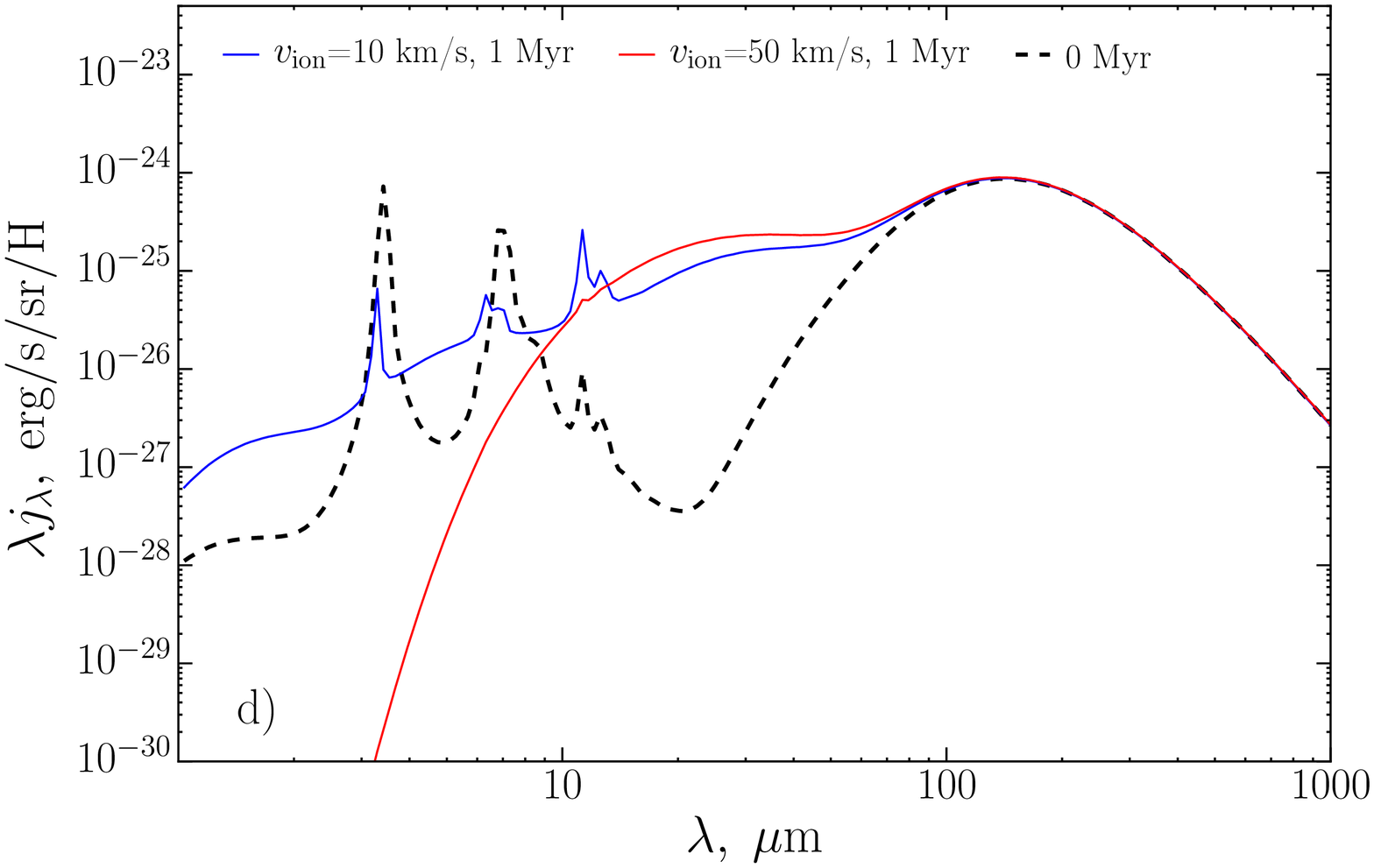}
\setcaptionmargin{5mm}
\onelinecaptionsfalse
\caption{Top left: dust emission spectrum at the initial time in the presence of
radiation with an intensity of $U=0.1$ (dashed curve) and $U=10$ (dotted curve),
and after one million years for $U=0.1$, 1, 10, 100, and 1000, $v_{\textrm{ion}}=10$~km/s and $E_0=2.9$~eV. Top
right: same for $E_0=5$~eV. Bottom left: dust emission spectrum at the initial time
(dashed curve) and after one million years for velocities of 
$v_{\textrm{ion}}=10$ and 50~km/s, with $U=1$ and $E_0=2.9$~eV. 
Lower right: same for $E_0=5$~eV.}
\end{figure*}

\section{EVOLUTIONARY VARIATIONS OF THE SPECTRAL AND PHOTOMETRIC 
CHARACTERISTICS OF A DUSTY MEDIUM}

We discussed above how the grain size and aromatization degree distributions
can change due to evolutionary processes. However, observations do
not allow directly judging the composition and 
distributions of dust over various parameters. Thus, it is necessary to
consider how evolutionary processes influence observed dust spectra. As was
noted above, we used the J13 distribution as the initial size distribution 
for the grains (Fig.~3a, black dotted curve), which satisfies the main
observational constraints (IR emission, absorption properties, particularly
in the UV, etc.).

Figure~5 shows dust emission spectra at the initial computation time and after one
million years for several sets of external conditions. Three parameters were
varied: the intensity of the radiation field $U$, the velocity of collisions
between the gas and dust $v_{\textrm{ion}}$, and $E_0$. The upper panels
show how the dust spectrum changes under the action of external radiation
fields of various intensities, for a fixed collision velocity of 10~km/s. 
The parameter $E_0=2.9$~eV for the plots in Fig.~5a and $E_0=5$~eV in Fig.~5b.
On all the panels, spectra for the initial size distribution are shown in
black, with $U=0.1$ (dashed) and $U=10$ (dotted). They characterize the 
emission of aliphatic hydrocarbons with strong bands at 3.4 and 7~$\mu$m.
The J13 initial distribution has a large quantity of small grains, which are
responsible for radiation at 2--20~$\mu$m, making these bands very strong.

After one million years with $E_0=2.9$~eV (Fig.~5a), even a weak
UV radiation field destroys most small grains, and only
the features at 11.3 and 12.7~$\mu$m, which are generated by aliphatic 
grains, remain in the near-IR spectrum. Emission bands corresponding to 
aromatic compounds are preserved only in the case of higher values of $E_0$
parameter, 5~eV (Fig.~5b). Thus, since aromatic bands at wavelengths shorter than 10~$\mu$m are in fact observed
in the ISM, a higher value of $E_0$ seems to be preferred. Variations in the
intensity of the radiation field $U$ influence the form of the continuum
(in particular, the position of the long-wavelength
maximum) only at wavelengths longer than 10~$\mu$m (in the considered range).

\begin{figure*}[t!]
\includegraphics[width=0.8\textwidth]{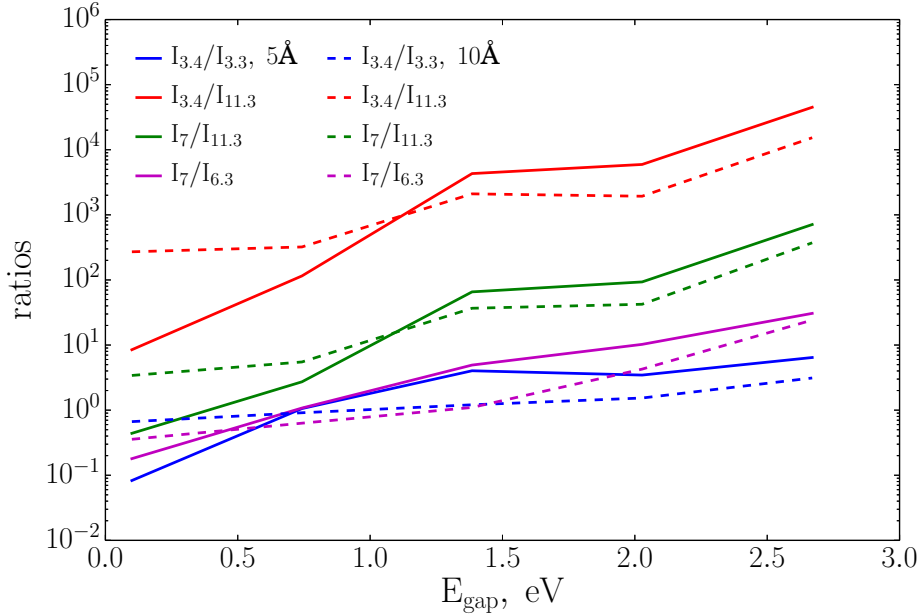}
\setcaptionmargin{5mm}
\onelinecaptionsfalse
\caption{Radiation intensity ratios in some aromatic and aliphatic bands for
grains with radii of 5~\AA\ (solid curves) and 10~\AA\ (dashed curves) for
$U=1$.}
\end{figure*}

Aromatization is efficient even when $U=0.1$. The computations show that the
aromatization rate is appreciably reduced only when $U=0.01$. Such low radiation
intensities can be encountered at the peripheries of galaxies, far from stars and
SFRs; however, all dust grains with radii smaller than ${\sim}100$~\AA\ in the
vicinities of stars, especially massive stars, will be aromatized on short 
times scales. Note that the radiation field plays a double role in the formation
of spectra, simultaneously disrupting small grains and causing them to 
radiate, so that the response of the spectra to an increase in the radiation
field is not linear.

The lower panels in Fig.~5 show the computed dust spectra for $U=1$ after
a million years of evolution and various velocities $v_{\textrm{ion}}$.  Also, as in
the upper panels, the black dashed curve shows the spectrum for the initial
size distribution. There are no aromatic bands in the spectrum when
$E_0=2.9$~eV and $v_{\textrm{ion}}=10$~km/s, but there is an appreciable
continuum at wavelengths of several microns, indicating the presence of
stochastically heated small grains. Appreciable aromatic bands also appear
in the spectrum when $E_0=5$~eV. Increasing the velocity to 50~km/s has
substantial consequences: bands in the near IR are not observed for
either $E_0$ value. This means that, in the presence of such velocities,
the rate of destruction of small grains exceeds the rate of their formation.

Thus, we can draw the following conclusions. (1) The intensity of the radiation
field is important for aromatization of grains, but, even when $U=1000$, in the absence
of IR bands, a substantial continuum is observed at wavelengths of several microns.
This means that the smallest aromatic grains are destroyed in this
medium, but stochastically heated grains smaller than 250~\AA\ in size are
preserved. The efficient destruction of any small grains is the result of
either collisions with gas particles with velocities above several tens of 
km/s or the presence of a very strong radiation field. (2) A C--C binding
energy of about 3~eV would lead to the rapid destruction of small grains and 
the disappearance of aromatic bands. Since such bands are observed, the
energy $E_0$ must be of the order of 5~eV or higher.

Since substantially less spectral data than photometric data on dust emission
in the ISM are available, it is interesting to trace how various external
conditions are reflected not only in the appearance of the spectrum, but also in
ratios  of integrated intensities in the near, middle, and far IR. The ratio of
fluxes at 8 and 24~$\mu$m wavelengths, which correspond to the central 
wavelengths of the \emph{Spitzer} IRAC and MIPS filters, have often been 
used to characterize the abundances of small grains and PAHs [42]. The
\emph{Herschel} Space Telescope (PACS instrument) can be used to study
dust emission in bands at 70, 100, and 160~$\mu$m with comparable angular
resolution. We computed synthetic intensities in these filters taking into
account response functions of the IRAC, MIPS, and PACS 
instruments\footnote{The response functions were taken from
http://ssc.spitzer.caltech.edu and http://herschel.esac.eas.int.}. 
In addition to the intensities at 8, 24, 70, and 160~$\mu$m, we also calculated
how they are related to the intensities at wavelengths of 3.3, 3.4, 6.3, 7,
and 11.3~$\mu$m. The response functions for these bands were chosen so that
they were equal to unity if a feature was located inside the interval and 0
if it was located outside the interval. The $I_{6.3}/I_{70+160}$ and
$I_{3.3}/I_{70+160}$ ratios can be used to obtain information about variations in
the abundance of small aromatic dust relative to the abundance of large
dust grains, while the $I_{3.4}/I_{70+160}$ and $I_{7}/I_{70+160}$ ratios 
can be used to estimate the abundance of aliphatic dust. Furthermore, we
separately considered $q_{\textrm{PAH}}$, the mass fraction of small
aromatic dust grains with radii $a< 20$~\AA{} and $E_{\textrm{gap}}<1$~eV
in the total mass of dust~[10, 43]. We again emphasize that, although
the conventional notation used for these grains is ``PAH'' (polycyclic aromatic 
hydrocarbons), we interpret this quantity more broadly, as the fraction of 
any even partially aromatized small grains.

\begin{figure*}[t!]
\includegraphics[width=0.4\textwidth]{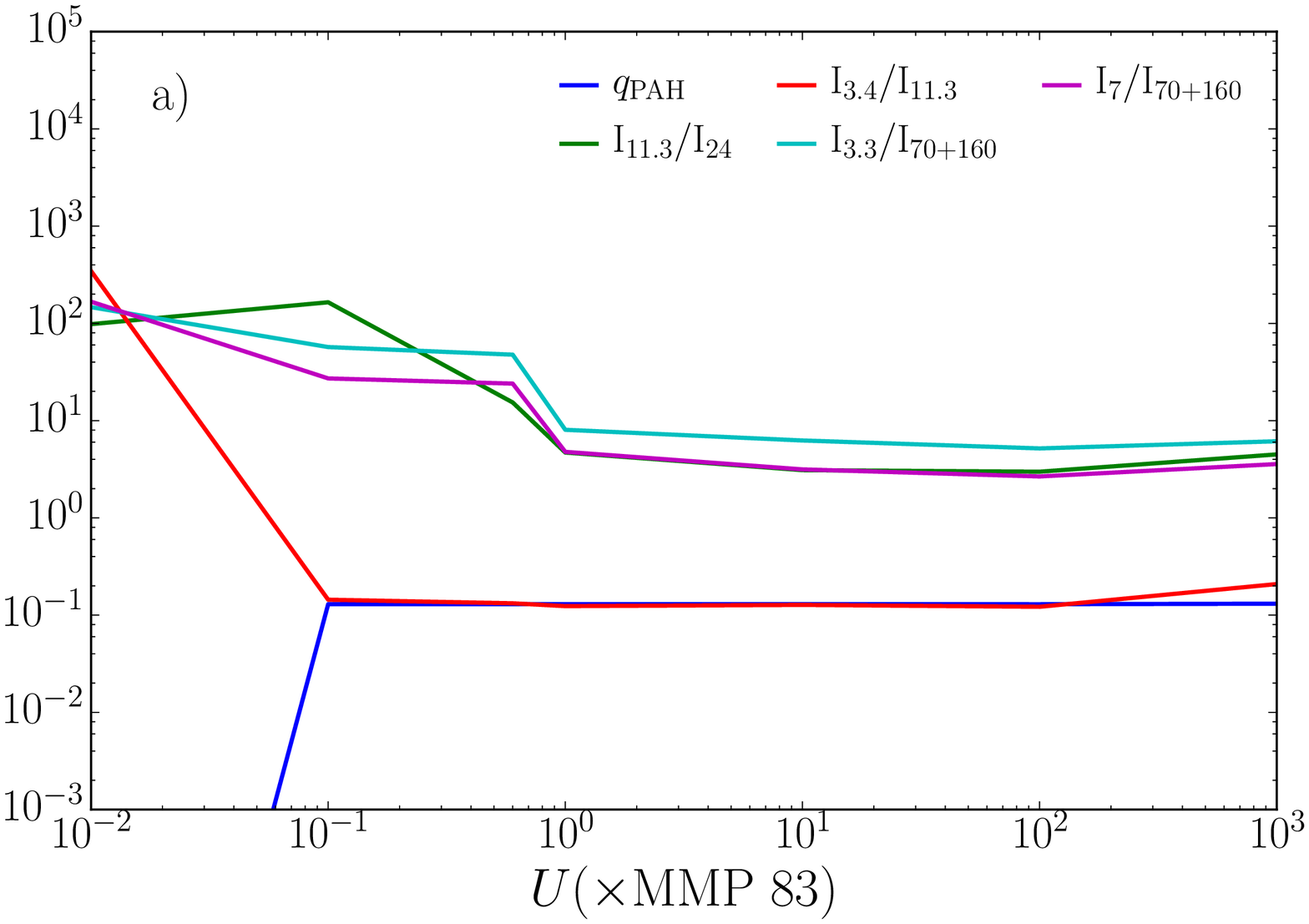}
\includegraphics[width=0.4\textwidth]{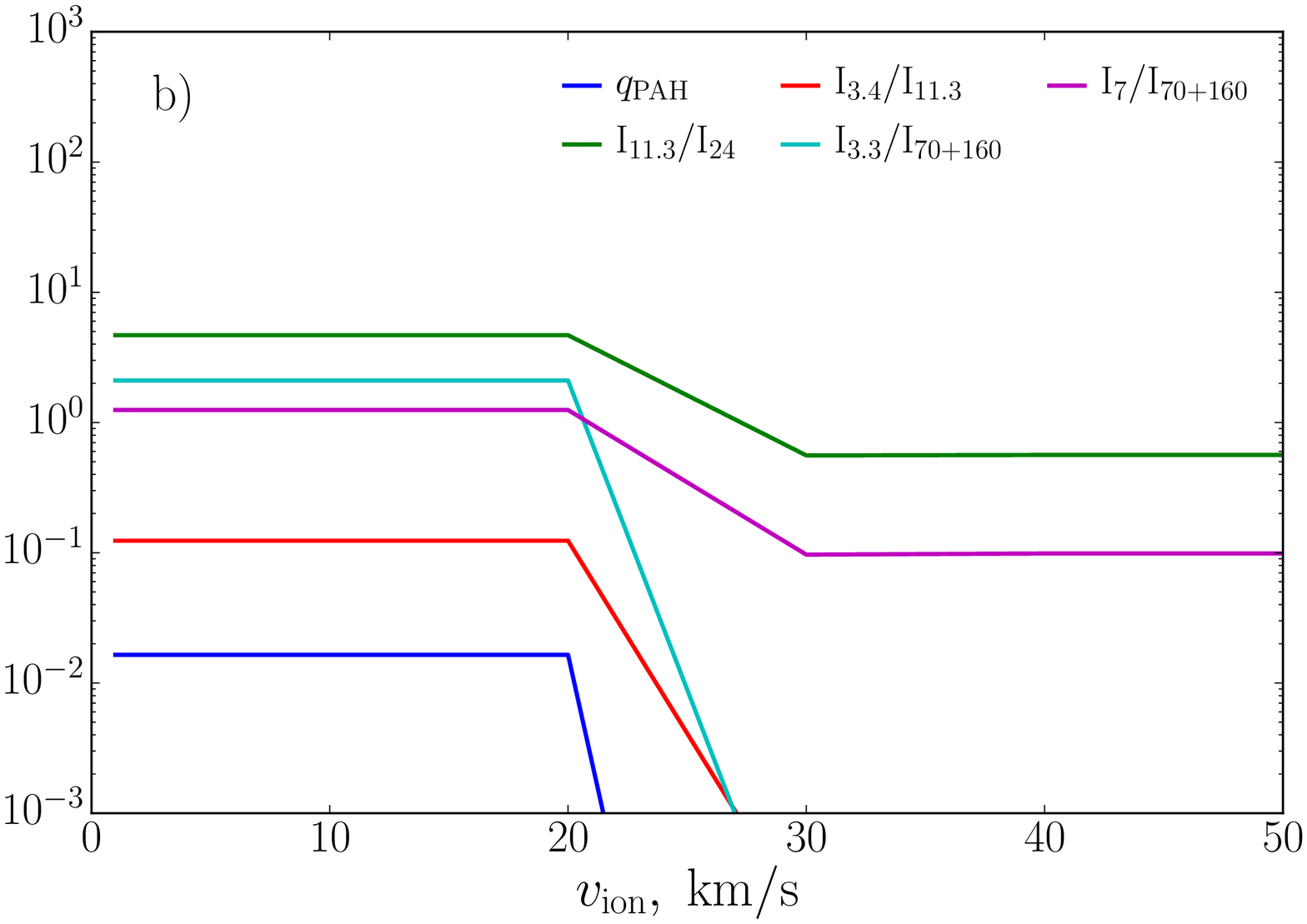}
\setcaptionmargin{5mm}
\onelinecaptionsfalse
\caption{Left: intensity ratios for various wavelengths and mass 
fractions of small aromatic grains as a function of the
radiation field intensity $U$ after one million years of evolution. Right: same, as 
a function of the velocity of collisions between gas and dust particles 
for $E_0=5.0$~eV.}
\end{figure*}

Figure~6 shows the intensity ratios for single grains with radii of
5 and 10~\AA\ for $U=1$ in some aromatic and aliphatic bands as a function
of the aromatization degree, expressed through the parameter
$E_{\textrm{gap}}$. The $I_{3.4}/I_{11.3}$ (red lines) and $I_{7}/I_{11.3}$
(green lines) ratios are most sensitive to $E_{\textrm{gap}}$, and are
especially high for the 5~\AA\ grains. The $I_{3.4}/I_{3.3}$ ratio for the 
5~\AA\ grains depends only weakly on $E_{\textrm{gap}}$, and this dependence
essentially disappears for the  $I_{3.4}/I_{3.3}$ ratio for the 10~\AA\
grains.

Further, we consider how various intensity ratios for an ensemble of grains
corresponding to the J13 initial size distribution depend on external
factors.

The dependences of the intensity ratios at different wavelengths after one
million years on the UV intensity and the relative velocity of collisions
between ions and dust grains $v_{\textrm{ion}}$ for $E_0=5$~eV are shown in 
Fig.~7. The $I_{3.3}/I_{70+160}$ ratio can be used to trace variations
in the radiation of small aromatic compounds relative to radiation of large 
grains. This ratio drops at $U<1$, then remains roughly constant. Figure~5a
shows that the intensity in the 3.3~$\mu$m band remains constant as $U$
increases, while the emission at 70 and 160~$\mu$m varies. When the intensity
of the radiation field increases, large dust grains are heated to higher
temperatures, and the emission maximum is shifted from longer to shorter
wavelengths. Accordingly, we first observe an increase in the intensity at
70 and 160~$\mu$m, after which this intensity remains constant. The ratio
$I_{3.3}/I_{70+160}$ is also non-zero when $U=0.01$, but, in this case, we
see emission of aliphatic rather than aromatic compounds, since the
band at 3.4~$\mu$m is nearby and fairly broad. A similar picture is observed
for the $I_{7}/I_{70+160}$ ratio. Aliphatic compounds are responsible for
the emission at 7~$\mu$m, but aromatic grains also have a band near this
wavelength; therefore, generally speaking, it is not possible to distinguish
one from the other for very low values of $U$, as in the previous case.
However, as a result of the evolution, small aliphatic grains are absent for
almost any $U$ value, so that the emission near 7~$\mu$m usually reflects the
composition of aromatic grains.

The $I_{3.4}/I_{11.3}$ ratio is the best tracer of the relative
abundance of aliphatic and aromatic grains. This ratio is very high at low
$U$ values, when all the grains are aliphatic, since the 11.3~$\mu$m band
is weak while the 3.4~$\mu$m band is strong. This ratio then falls when
aromatic grains appear in the medium, but not as much as it would in the 
absence of a contribution from the wings of the 3.3~$\mu$m band. Thus, good
quality spectra are essential for accurately separating the bands due
to the different types of grains.

We also considered the dependence of the $I_{11.3}/I_{24}$ ratio on the UV
radiation field. This ratio reaches a maximum at $U\approx0.1$, and falls
only slightly at lower values of $U$, since the aliphatic 11.3~$\mu$m band 
is very weak. This decrease at small $U$ values is insignificant, since the
24~$\mu$m emission also decreases with the intensity of the radiation field.
The $I_{11.3}/I_{24}$ ratio drops stronger when $U$ is increased, due to
an appreciable increase in the 24~$\mu$m emission. A modest growth is again
observed for high radiation fields ($U>100$), due to the growth in the 
11.3~$\mu$m-band intensity.

The blue curve in Fig.~7 shows the dependence of the mass fraction of small
aromatic grains $q_{\textrm{PAH}}$ on $U$. When $U<0.1$, $q_{\textrm{PAH}}$
is modest, since the aromatization rate is low in the presence of such weak
radiation, and not even the smallest grains are able to be constructed. When
the radiation field is increased to $U=0.1$ or more after one million years,
all small grains with sizes corresponding to PAHs become aromatic. In the
presence of stronger radiation fields, this fraction may again decrease due
to the destruction of small aromatic grains by UV radiation, but such strong
fields can exist only in the immediate vicinities of hot stars.

Figure~7b shows how these same ratios vary with the collisional velocity. 
It is obvious that they are much more sensitive to this parameter.  When the
collisional velocity between the ions and dust grains is 30~km/s, the fraction
of small aromatic grains ($q_{\textrm{PAH}}$) becomes negligible after one
million years. The 3.3 and 3.4~$\mu$m emission bands are essentially
absent by this time, causing the $I_{3.3}/I_{70+160}$ ratio to similarly drop 
sharply at velocities above 20~km/s. The $I_{7}/I_{70+160}$ and 
$I_{11.3}/I_{24}$ ratios decrease by an order of magnitude, but not as
sharply as the $I_{3.3}/I_{70+160}$ ratio. This occurs because some 
contribution from large grains is possible both at 7 and 11.3~$\mu$m, while
this is not the case at 3.3 and 3.4~$\mu$m.

In this study, we have considered only the emission properties of hydrocarbon
grains. The shape of the continuum in the region of the considered
aromatic and aliphatic
bands can also be influenced by the presence of silicate absorption 
bands, which slightly change the considered ratios quantitatively.  However, 
the qualitative behavior of these ratios remains the same. The detailed
analysis of galaxy spectra from the SINGS survey presented in [44] did not
detect significant indications of the presence of silicate absorption bands in
nearly $90\%$ of the studied galaxies. Our additional spectral computations
show that these features begin to show up in emission only when
$U=10^4{-}10^6$. This is consistent with the computational results of~[9].

\section{CONCLUSION}

The presented results show that the evolution of an ensemble of
dust particles strongly depends on the initial distributions of the dust size
and aromatization degree. The J13 and MRN initial distributions
that we have considered yield different results. The evolution of the J13
distribution, in which the mass fraction of small grains is initially large,
is appreciably determined by the relative velocity of collisions with gas 
particles and the radiation field. In contrast, shattering processes play a 
key role in the MRN distribution, substantially reducing the fraction of large
grains and increasing the mass fraction of small grains with time, partially
compensating the influence of other factors that destroy small grains. The
destruction of small grains by gas particles and radiation plays a ``positive''
role in preserving large grains in the J13 model, since one important factor
leading to the destruction of large grains is shattering during collisions
with smaller grains.

In the range of $U$ values considered, the radiation field chiefly leads to
grain aromatization, while efficiently destroying only the smallest grains.
Sputtering due to collisions with gas particles leads to the destruction of
larger grains. Grains with sizes below 20~\AA\ are absent when the relative
collision velocities are higher than 50~km/s. The destruction of grains
will be even more efficient in the presence of higher velocities, such as
are characteristic of supernovae.

The IR emission spectrum of dust depends substantially on its evolution.
The shape of the near-IR (2--15~$\mu$m) spectrum of evolved dust in the
case of high $E_{0}$ values is virtually independent of the intensity of
the radiation field for $U$ values from 0.1 to $10^4$. Aromatization is
not efficient at lower $U$ values, and characteristic aromatic bands are
absent from spectra in this case. As the radiation field is increased,
we observe a transition to spectra containing features characteristic for
aromatic grains. The destruction of small grains is especially efficient
for C--C bond energies $E_0=2.9$~eV. Even in the case of parameters of the
medium characteristic of the ``ordinary'' ISM, small aromatic particles 
are completely absent in a medium with this $E_0$ value after one million 
years of evolution, leading to the absence of aromatic bands in its spectrum.
Since these bands are preserved only when $E_0$ is about 5~eV, this indicates
a preference for relatively high values of $E_0$. 

The $I_{3.4}/I_{11.3}$ and $I_{7}/I_{11.3}$ intensity ratios demonstrate the
greatest sensitivity to the aromatization degree of small grains. The
$I_{3.4}/I_{3.3}$ ratio, which has often been proposed as a measure of the
aromatization of small grains, is virtually independent of $E_{\textrm{gap}}$
for grains with radii of 5--10~\AA. The $I_{3.4}/I_{11.3}$ ratio is a convenient 
indicator of the relative abundances of aliphatic and aromatic grains for 
various values of $U$ and $v_{\textrm{ion}}$. The $I_{3.3}/I_{70+160}$ ratio 
is a sensitive indicator of the contribution of aromatic grains to the total 
mass fraction of dust. The 3.3 and 3.4~$\mu$m bands 
partially overlap, so that distinguishing each of them clearly requires 
good-quality observational data.

In this study, we considered only conditions that do not differ appreciably
from those in the thermal phase of the ISM. It is obvious that the destruction
of grains will be more efficient under the more extreme conditions inside 
regions of ionized hydrogen and in supernova remnants, which should be reflected
in the corresponding dust emission spectra. Moreover, the spectra of the
external radiation in such regions could differ from the spectrum considered
in~[21]. Shattering during high-velocity collisions between grains may
not compensate for the destruction of small grains by the radiation field in this
case, so that the corresponding spectra would differ from those presented here.

Our model can be used to consider various situations, including some not
considered here. Its further application and optimization (especially the
inclusion of grain charge, consideration of the evolution of silicate grains,
and 3D computations) will be considered in future. The model
can also be applied to studies of specific objects, such as IR bubbles around
HII reagions, planetary nebulae, and supernova remnants.

\section{ACKNOWLEDGEMENTS}

We thank the referee for important comments, and V. Akimkin and A. Jones for
useful discussions. This work was supported by the Russian Foundation for
Basic Research (grants 14-02-00604, 14-02-31456, 15-02-06204), a President
of the Russian Federation Grant (MK-4536.2015.2), and the ``Dinastiya''
foundation.

\end{document}